\documentclass[conference]{IEEEtran}
\IEEEoverridecommandlockouts
\usepackage{cite}
\usepackage{comment}
\usepackage{amsmath,amssymb,amsfonts}
\usepackage{algorithmic}
\usepackage{graphicx}
\usepackage{gensymb}
\usepackage{textcomp}
\usepackage{xcolor}
\usepackage{makecell}
\usepackage{multirow}
\usepackage{subfigure}
\def\BibTeX{{\rm B\kern-.05em{\sc i\kern-.025em b}\kern-.08em
    T\kern-.1667em\lower.7ex\hbox{E}\kern-.125emX}}
\begin{document}

\title{
An Optimal Baseband Delay-Based Beam Squint Removal Scheme across a Range of Steering Angles for Digital Wideband Beamformers in Radars}
\author{\IEEEauthorblockN{\textsuperscript{1}Neeraja P. K., \textit{Graduate Student Member, IEEE},
\textsuperscript{2}Bindiya T. S., \textit{Senior Member, IEEE}, \\ \textsuperscript{3}Raghu C.V., \textit{Member, IEEE}}

\IEEEauthorblockA{\textsuperscript{1,2,3}\textit{Dept. of Electronics and Communication Engg., National Institute of Technology Calicut, Kerala, India}} 
\textsuperscript{1}neerajapkn@gmail.com,
\textsuperscript{2}bindiyajayakumar@nitc.ac.in,
\textsuperscript{3}raghucv@nitc.ac.in
\vspace{-0.3cm}}

\maketitle
\thispagestyle{plain}
\pagestyle{plain}
\pagenumbering{gobble} 
\begin{abstract}
This paper is an attempt to mitigate the beam squint happening due to frequency-dependent phase shifts in the wideband beamforming scenario, specifically in radar applications. The estimation of the direction of arrival is significant for precise target detection in radars. The undesirable beam squint effect due to the phase shift-only mechanism in conventional phased array systems, which becomes exacerbated when dealing with wide bandwidth signals, is analyzed for a large set of steering angles in this paper. An optimum baseband delay combined with the phase shift technique is proposed for wideband radar beamforming to mitigate beam squint effectively.
This technique has been demonstrated to function properly with 1-GHz carrier frequency for signals with wide bandwidths of up to $\pm250$MHz and for steering angles ranging from 0 to 90 degrees.
\end{abstract}


\section{Introduction}
Radar systems transmit signals in a range of bandwidths, typically spanning from a few megahertz to several hundred megahertz, to carry out a diverse range of tasks such as target detection, tracking, imaging, and identification.
%
Beamforming plays a crucial role in electronically steering the beam, allowing the phased array system to rapidly change the main shaft without the need for physical movement of the antenna array compared to traditional mechanical steering of beams. Through beamforming, which is spatial filtering, a beam with maximum power is formed where the target is located. Steering to the correct angle depends on various factors such as antenna array factor, weighting vector, etc.



Narrowband beamforming is tailored for narrowband frequency range and gives high-resolution beam steering, whereas wideband beamforming considers wide frequency range, making it more versatile \cite{art1}.  However, since most of the earlier wireless communication applications primarily deal with signal bandwidth which is relatively narrow, almost all of the earlier antenna literature is focused on narrowband beamforming \cite{art2} -\cite{art3}. The ability to process a wide range of frequencies enables better discrimination between closely spaced targets, and hence wideband beamforming is important in diverse areas ranging from military and air defense, ground surveillance, missile tracking, radio astronomy, and even seismology, medical diagnosis and treatment, to communications. 
It is also used in weather radar for studying severe weather phenomena and in civil aviation for air traffic control and weather monitoring. Such a weather radar system was recently developed using the multiple-input multiple-output (MIMO) technology in \cite{art4}. 


%
In the case of wideband beamforming, frequency-dependent phase shifts alone cannot provide the required time delay because the frequencies involved are a range of frequencies \cite{art5}. The steering weight vector corresponding to a single frequency will not help to steer the beam correctly, and beams will be shifted to the left or right of the desired steering angle; this phenomenon is termed beam squint \cite{art1}. The paper \cite{art6} is one of the fine articles on phased antenna arrays, and it gives a crisp understanding of the beam squint problem in wideband arrays. 
The changes in the signal's frequency cause changes in the phase relationships between the elements, which causes the primary beam to move. This sensitivity may impose challenges in applications such as target recognition and tracking, where a stable and precisely directed beam is necessary. 

Recently, many works have been reported that deal with the techniques to mitigate the beam squint in wideband antenna arrays \cite{art7} -\cite{art16}.
To reduce the beam squint effects in the low earth orbit satellite integrated sensing and communications system, the authors of \cite{art7} proposed an algorithmic method that uses statistical channel state information knowledge for hybrid precoding. It considers statistical wave propagation properties and characterizes beam squint effects, providing insights into efficiently utilising spectral and hardware resources in the space-air-ground-sea integrated network. The use of abnormal group delay phase shifters to remove beam squint in series-fed array antennas is discussed in \cite{art8}. However, they may introduce a nonlinear phase response across the frequency band, which could result in distortions in the antenna pattern and sidelobes, affecting the overall performance of the array.
To overcome these challenges and offer a more robust solution, the true time delay (TTD) technique emerges as a compelling alternative \cite{art9}, \cite{art10}. 
Unlike the other methods, TTD offers several distinct advantages. One key benefit is its ability to introduce time delays to the signals rather than relying solely on phase shifts, thereby minimizing nonlinear phase responses that can lead to pattern distortions. This characteristic of TTD ensures a more precise and controlled adjustment of the antenna beams.
The surge in the number of works utilizing TTD techniques for beamforming, as evidenced by recent research \cite{art9}-\cite{art16}, signifies a growing recognition of TTD's efficacy. 
One notable approach involves the use of radio frequency (RF) timed arrays, as discussed in \cite{art11}, to introduce RF time delays aimed at mitigating array intersymbol interference (ISI) and squint concerns. 
Despite their effectiveness in handling certain aspects of beam squint, these arrays face challenges such as a single-beam limitation, high power consumption, and substantial area demands. It is also important to note that RF time delay and baseband time delay are distinct.
In the pursuit of effective solutions for mitigating beam squint in wireless communication scenarios, the work presented in \cite{art14} adopts a distinctive baseband TTD approach. The methodology employed involves a meticulous process, wherein the initial time-delayed received signal is subjected to a series of operations aimed at achieving a squint-free signal. Specifically, the technique in \cite{art14} leverages the multiplication of the initial time-delayed received signal by a delayed baseband signal. This multiplication process is instrumental in shaping the characteristics of the received signal, setting the stage for subsequent phase-shifting operations. 

This work provides a strategy for reducing beam squint problems when dealing with wide bandwidth signals in radars by combining a finely adjustable fractional baseband delay with phase shift. The contributions of this work are briefly summarized as follows:
\begin{itemize}
	\item An extensive study of the beam squint errors in the phase shift-only antenna arrays when dealing with wideband signals is included in this paper.
    \item A joint method of fine baseband delaying with phase shifting technique is employed for the radar wideband beamforming scenario to mitigate the beam squint problem that occurs due to narrow-band assumption in the phased arrays.
    \item An optimization problem is formulated to find the delay for the entire wide bandwidth range that could effectively reduce the  beam squint at each steering angle.
    \item Detailed analysis of beam squint removal while using the proposed technique over a wide range of bandwidths and steering angles is performed.
\end{itemize}

The rest of this paper is structured as follows. The proposed method for the squint mitigation scheme based on the optimum baseband delay combined with the phase shift is explained in Section II. The design example in Section III provides the simulation results to illustrate the effectiveness of the proposed technique in squint removal in wideband antenna arrays. Section IV concludes this paper. 
 
\section{Proposed Optimal Baseband Delay-Based Beam Squint Removal Scheme for Radar Wideband Beamformers}
Issues due to the narrow band assumption while dealing with wideband signals and the proposed method to solve the problem are explained in this section. 
For narrow-band incoming signals, the time delay to steer the beam in the correct direction is achieved by employing a phase shift in the phased array architecture. The relationship between phase shift and time is given in the equation below\cite{art9}:
\begin{equation}
\small
    \mathrm{Phase\ Shift(rad)}=2\pi \times \mathrm{Frequency(Hz)} \times \mathrm{Time\ Delay(sec)}
\end{equation}
This phase shift ensures that when the signals from multiple elements are combined, they constructively interfere in the desired direction and destructively interfere in other directions, creating a beam of high gain in the desired direction. 

\begin{figure}[!htbp]
\centerline{\includegraphics [width=8.5cm,height=3.5cm]{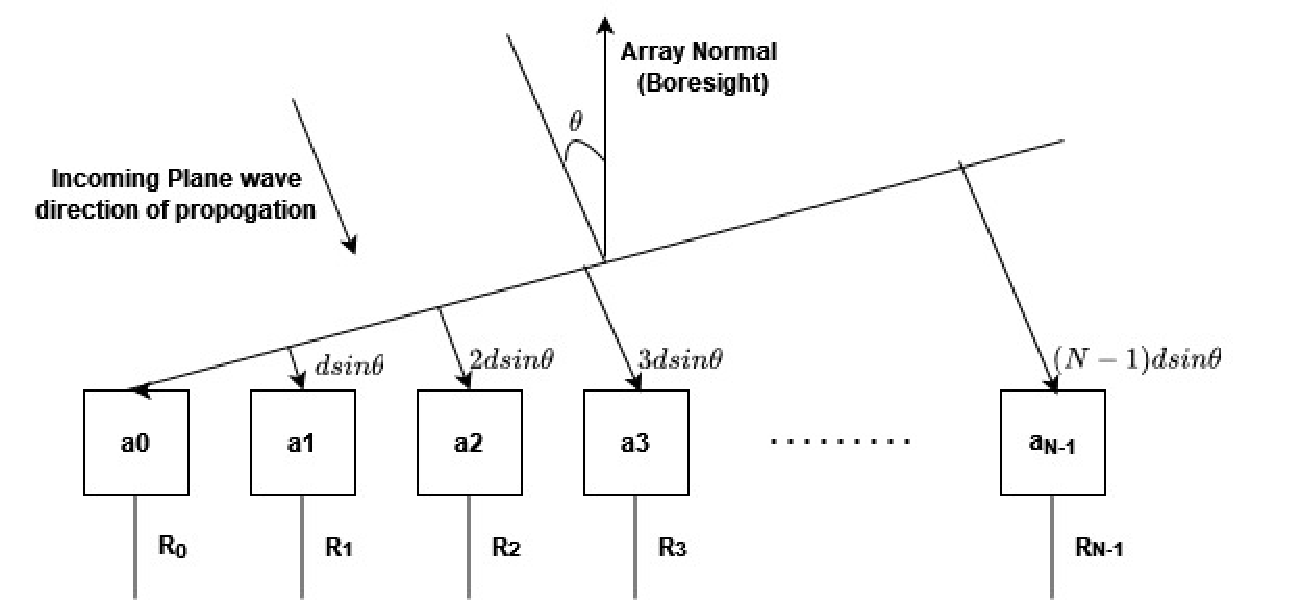}}
\caption{ULA configuration of phased array radar receiver }
\label{ULA}
\end{figure}

But when the conventional phase shift beamformer is used for wideband signals, beam squint occurs \cite{art9}.
Hence, as a solution to the squint problem,  a time delay-based technique along with fine-tuned phase shift is discussed in \cite{art14} for wireless communication applications. In our work, this method is modified for wideband radar applications using a combination of integer baseband delay, fine-tuned fractional baseband delay, and a constant phase shift to precisely steer the beam when dealing with wideband signals. 
To understand the need for combining baseband delay and phase shifting to eliminate the squint, a brief explanation is given below:

Let us consider a uniform linear array (ULA) consisting of the $N$ number of array elements with an inter-element spacing of $d=\lambda /2$ to avoid spatial aliasing, where, $\lambda$ is the wavelength and the carrier frequency be represented as $f_c$.  Let us consider a plane wave impinging on the ULA at an angle $\theta$ as shown in Fig. \ref{ULA}. The first antenna element is denoted as $a_{0}$,  the second element is denoted as $a_{1}$ and so on. The received signals at individual antenna elements are denoted by $R_{0}$, $R_{1}$ etc. Let the baseband signal, $b(t)$, be an exponential wave represented as,
\begin{equation}
b(t) = e^{j2\pi f_{bb}t}
\end{equation}
where, $f_{bb}$ is the baseband frequency. Assuming a carrier with frequency $f_c$ is used for modulation, the modulated signal can be represented as
\begin{equation}
m(t) = e^{j2\pi {(f_c \pm f_{bb})t}}
\end{equation}

The propagation delay incurred between the signal received at the first element and the $n^{th}$ element in the receiver array is given by \cite{art14}
\begin{align}
\tau_{n} = &(n-1)d\ \frac{\sin\ \theta }{c}, \qquad \qquad            n = 1, 2, 3, 4,..., N
\label{element delay}         
\end{align} 
where $c = 3\times10^{8}$ m/s is the speed of light in free space.
Thus, the signal received at the $n^{th}$ element in the antenna array is given by
\begin{equation}
\begin{aligned}
r(t) &= e^{j2\pi {(f_c \pm f_{bb})(t-\tau_n)}}\\
 &= e^{j2\pi {(f_c \pm f_{bb})t-{f_c \tau_n}\mp {f_{bb} \tau_n}}}
\label{recsig}
\end{aligned}
\end{equation}
In the case of narrowband signals, since $f_c \pm f_{bb} = f_c$, the dependency on the propagation delay can be eliminated, and the original signal $m(t)$ can be restored from all the array elements by providing a phase shift equal to ${f_c \tau_n}$ to $r(t)$, the signal received. However, this assumption becomes invalid for wide-bandwidth signals and causes beam squint errors when phased arrays are employed in wideband scenarios. The beam squinting error in the phased arrays can be estimated as \cite{art17}
\begin{equation}
\begin{aligned}
\Delta\phi &= -\frac{\Delta f}{f_c}\tan\ \theta \\
&= -\frac{ \pm f_{bb}}{f_c}\tan\ \theta
\end{aligned}
\label{beamsquint}
\end{equation}
where $\Delta\phi$ is the beam squint, $\theta$ is the steered angle, $f_c$ is the carrier frequency, and $\Delta f$ is the offset frequency. 
It can be seen from Eqn. \eqref{recsig} that if in addition to the phase-shift of $f_{c}\tau_{n}$, a baseband delay of $\tau_{n}$ is also included for wide bandwidth signals, the original signal can be correctly retrieved, eliminating the squint. 
Thus, it is concluded that the received signal can be multiplied by the delayed baseband and then phase shifted to get the squint-free response when dealing with wideband signals. 

The baseband delays are often fractional values, which can be realized using a fractional delay filter along with an integer delay filter. However, since the baseband signal frequency varies, the delay to be introduced also varies, and an optimum fractional delay needs to be found for each steering angle.
This is formulated as an optimization problem and a suitable objective function is found as discussed in the next section.

\subsection{Optimizing the Baseband Fractional Delay and the Objective Function}
An appropriate amount of delay can compensate for the time difference between the signals reaching each antenna element. In the wide bandwidth scenarios as mentioned in the previous section, this delay is provided by a combination of baseband delay and a phase shift. However, the delay to be provided varies with the steering angles and baseband frequencies. In this work, this delay, which is a fractional delay, is proposed to be provided by a combination of an integer delay filter and a fractional delay filter. In practice, the main challenge in designing a fractional delay filter is selecting the appropriate number of taps (filter length) and the values of the coefficient multipliers to approximate the sinc function accurately. The choice of filter length affects the trade-off between the delay, accuracy and computational complexity. Longer filters can provide more accurate fractional delays but at the cost of increased computational requirements. Here, a fixed low-order FIR filter provides the integer part of the baseband delay for all the bandwidths under consideration across all the steering angles between $0^{\circ} - 90^{\circ}$. The problem of finding out the remaining fractional baseband delay to minimize the squint for each steering angle between $0^{\circ} - 90^{\circ}$ for the entire range of wide bandwidths is formulated as an optimization problem. This fractional delay can be achieved through fractional delay filters \cite{art18}.

Thus, the objective of the optimization problem is to find a delay to minimize the beam squint for a particular steering angle over the entire range of frequencies. The solution space includes the fractional delay values varied between $-f_s \times d/c$ to $f_s \times d/c$. For a particular steering angle, the beam squint error is calculated for each value of the fractional delay in the solution space for all the signal bandwidths under consideration. The objective is formulated to find a delay to minimize the sum of beam squint errors across all the signal bandwidths for each steering angle. Thus, the cost function or the objective function is defined as the total beam squint for all the signal bandwidths under consideration as given by
%
%
%
\begin{equation}
C  = \sum_{f_{bb}(i)}{|\theta'(f_{bb}(i)) - \theta|}, \ \ \mathrm{for}\ i=1,2,3,...,p
\label{opt_cost}
\end{equation}
where $f_{bb}(i)$ is the $i^{th}$ baseband signal frequency (out of a total of p) under consideration between the minimum and maximum baseband frequencies, $\theta'(f_{bb}(i))$ is the steered angle at the baseband frequency $f_{bb}(i)$, and $\theta$ is the required steering angle. Thus, the cost function is calculated for a set of fractional delay values in the solution space between $-f_s \times d/c$ to $f_s \times d/c$. The objective is to find an optimum delay value that minimizes the cost function given in Eqn. \ref{opt_cost}.

\begin{figure}[h]
\centerline{\includegraphics [width=8.5cm,height=4cm]{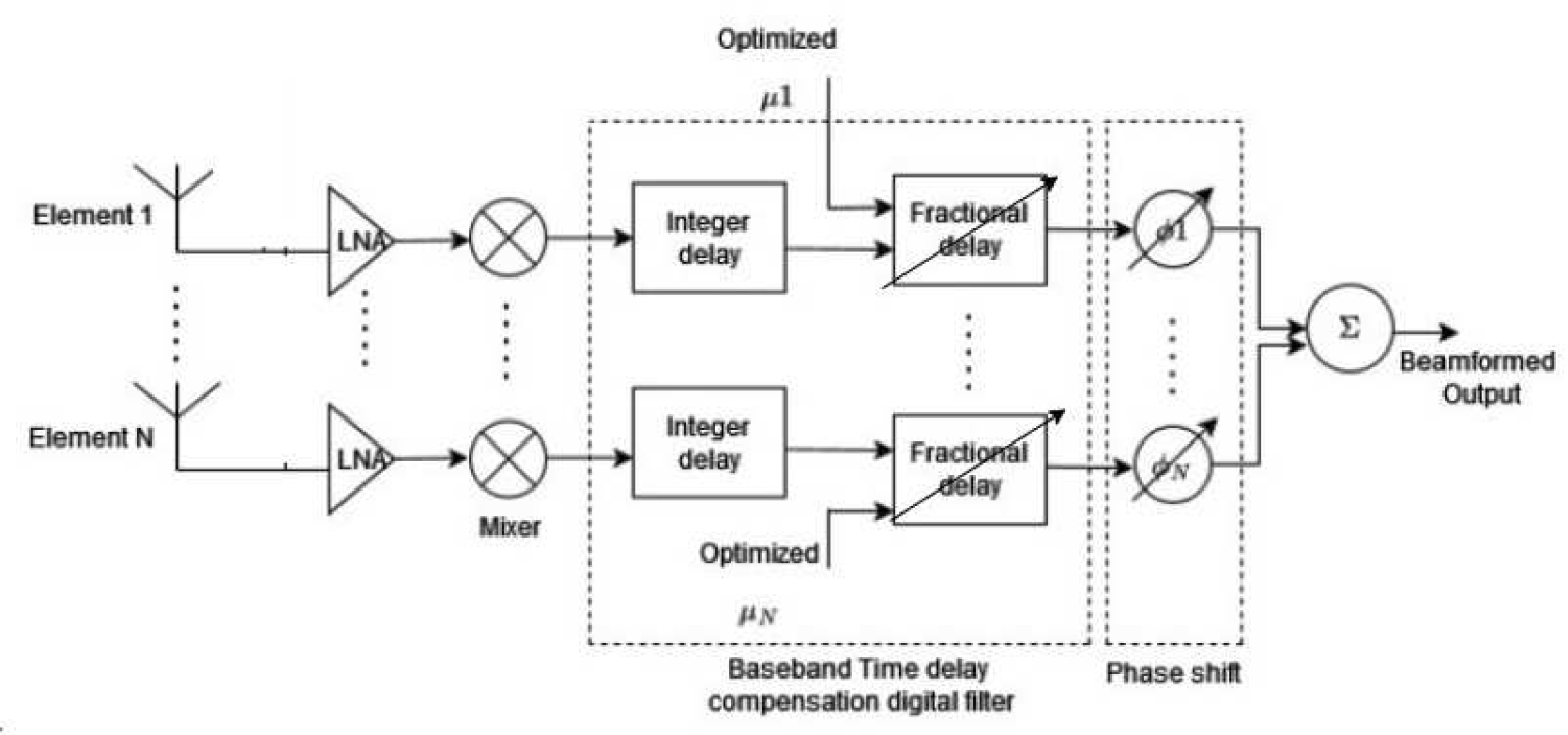}}
 \caption{Block diagram of the proposed optimized baseband delay-based beam squint elimination scheme in wideband radars}
 \label{fig4}
 \end{figure}

Thus, the solution to this optimization problem gives an optimum delay value that reduces the beam squint error at each steering angle across all the desired signal bandwidths. Thus, the optimum fractional delay value also changes as the steering angle changes. These delays can be used to design the variable fractional delay filter which can be reconfigured using the optimum delay values obtained w.r.t the steering angles. Thus, fine-tuning is introduced in the baseband delay that makes the true time delay precise. The proposed architecture is shown in Fig. \ref{fig4}.

Although the current work deals with element-wise digital beamforming, if the number of elements increases, the hardware resources and cost of the system will increase dramatically. It is beneficial most of the time for portable systems to digitize at the sub-array level using analog phase shifting at each array element. Such a phased array architecture using quadrant-wise digital channels has been demonstrated to be very cost-effective, powerful, and portable at X-band in \cite{art19} as a Portable Weather Radar (PWR). Their remarkable sub-array architecture elaborated in \cite{art4} and \cite{art19} enabled electronic beam switching capabilities in elevation while mechanically spinning in azimuth. Their approach resolved hardware complexity, which comes with an increasing number of digital channels with phased arrays, in turn helping to reduce cost and power budget and bring long-range sensing to smaller portable systems. The true time delay system evolved in our research is ideally suited for such platforms, and our future endeavours would be to bring true time delay to sub-array structures.

\section{Design Example}\label{AA}
This section gives a design example to analyze the suitability of the proposed technique of combining the optimal delay and phase shift for removing beam squint in the radar wideband beamformers over a range of steering angles. To demonstrate this, the steering angle is varied between $0^{\circ}$ to $90^{\circ}$ for a large range of wide bandwidths. The specifications of the radar beamformer are considered as follows:
\begin{itemize}
    \item Number of array elements: 16
    \item Carrier frequency: 1GHz
    \item Range of signal bandwidths: $\pm25$MHz - $\pm250$MHz
    \item Sample rate: 550MHz
    \item Range of steering angles: $0^{\circ} - 90^{\circ}$
    \item Beam Squint $< 1^{\circ}$
\end{itemize}

This section shows some interesting simulation results obtained. To illustrate the issue of beam squint for the wide bandwidth signals, initially, beamforming is performed using conventional phase array radar (phase-shift only). A set of 10 bandwidths in the desired range, $\pm$25MHz, $\pm$50MHz, $\pm$75MHz, $\pm$100MHz, $\pm$125MHz, $\pm$150MHz, $\pm$175MHz, $\pm$200MHz, $\pm$225MHz and $\pm$250MHz are considered for the analysis of the beam squint. The steering  angles considered are $0^{\circ}$, $10^{\circ}$, $20^{\circ}$, $30^{\circ}$, $40^{\circ}$, $50^{\circ}$, $60^{\circ}$, $70^{\circ}$, $80^{\circ}$ and $90^{\circ}$.

\begin{figure}[!htb]
	\centering
	\subfigure{\includegraphics[width=0.22\textwidth, keepaspectratio]{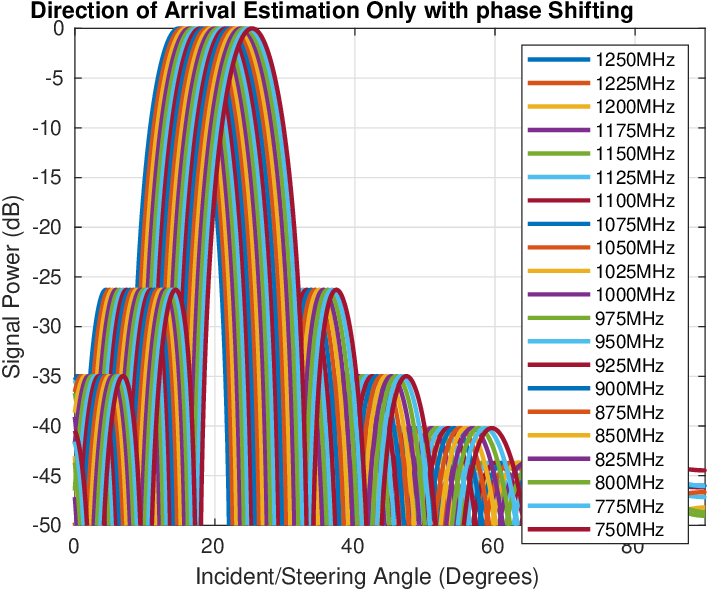}}
	\subfigure{\includegraphics[width=0.22\textwidth, keepaspectratio]{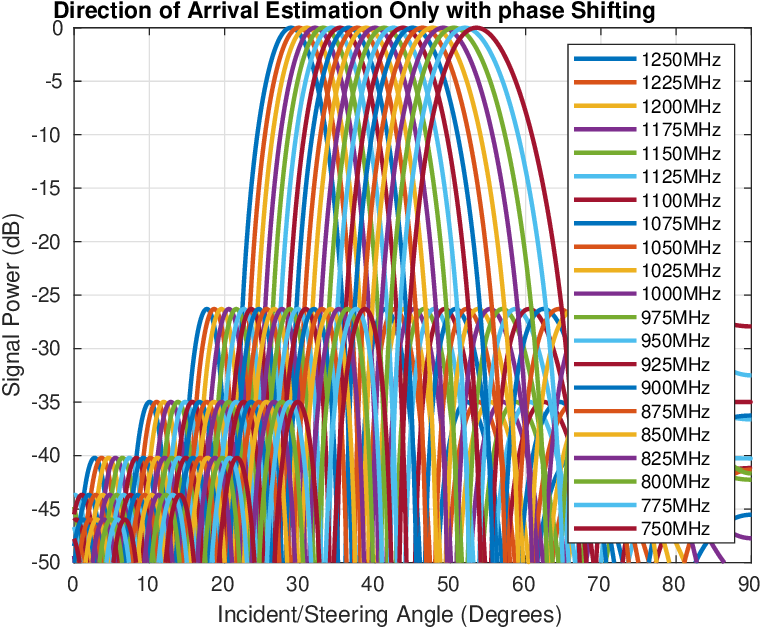}}
	\subfigure{\includegraphics[width=0.22\textwidth, keepaspectratio]{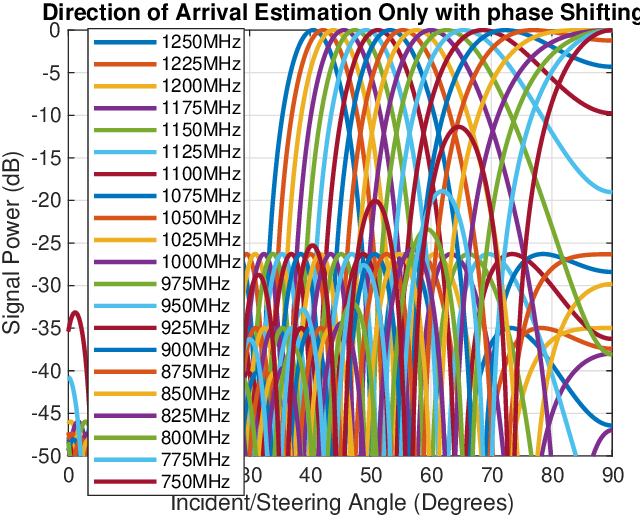}}
	\subfigure{\includegraphics[width=0.22\textwidth, keepaspectratio]{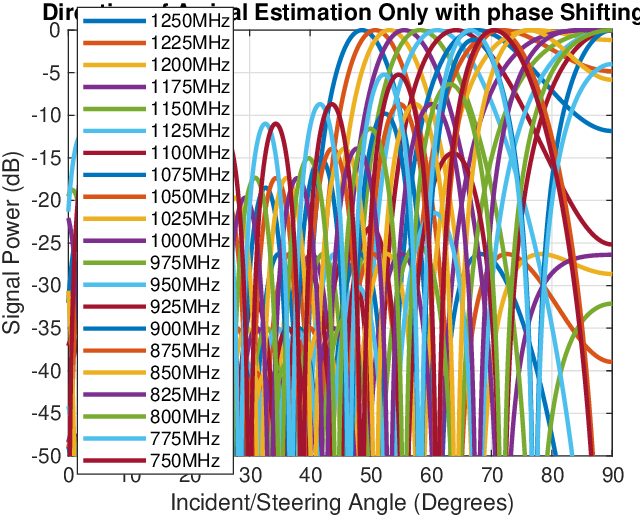}}
	\caption{ Beam squint error in phase array radars for different wide bandwidth signals when steering angle is (a) $20^{\circ}$ (b) $40^{\circ}$ (c) $60^{\circ}$(d) $90^{\circ}$}
	\label{beam_squint}%
\end{figure}

\begin{table}[!htb]
    \centering
    \caption{The beam squint for wide bandwidth signals for the steering angle of $40^\circ$ with conventional phase array beamformer}
    \resizebox{0.3\textwidth}{!}{
    \renewcommand{\arraystretch}{1}
    \begin{tabular}{|c|c|c|}
    \hline
         \makecell{Centre\\Frequencies} & \makecell{Steered\\Angle ($\theta'$)}  & \makecell{Beam Squint\\Error ($\Delta \phi$)}   \\ \hline
         750 MHz& $53.47^{\circ}$  & $13.47^{\circ}$\\
         \hline
         775 MHz& $51.94^{\circ}$  & $11.94^{\circ}$\\
         \hline    
         800 MHz& $50.48^{\circ}$  & $10.48^{\circ}$\\
         \hline
         825 MHz& $49.04^{\circ}$  & $9.04^{\circ}$\\
         \hline        
         850 MHz& $47.67^{\circ}$  & $7.67^{\circ}$\\
         \hline
         875 MHz& $46.32^{\circ}$  & $6.32^{\circ}$\\
         \hline         
         900 MHz& $44.99^{\circ}$ & $4.99^{\circ}$\\
         \hline
         925 MHz& $43.71^{\circ}$ & $3.71^{\circ}$\\
         \hline         
         950 MHz& $42.45^{\circ}$ & $2.45^{\circ}$\\
         \hline
         975 MHz& $41.21^{\circ}$ & $1.21^{\circ}$\\
         \hline         
         1000 MHz & $39.99^{\circ}$ & $0.01^{\circ}$\\ 
         \hline
         1025 MHz & $38.80^{\circ}$ & $1.20^{\circ}$\\ 
         \hline         
         1050 MHz& $37.63^{\circ}$ &  $2.37^{\circ}$\\
         \hline
         1075 MHz & $36.48^{\circ}$ & $3.52^{\circ}$\\ 
         \hline         
         1100 MHz & $35.36^{\circ}$ & $4.64^{\circ}$\\ 
         \hline
         1125 MHz & $34.23^{\circ}$ & $5.77^{\circ}$\\ 
         \hline         
         1150 MHz& $33.13^{\circ}$ &  $6.87^{\circ}$\\
         \hline
         1175 MHz & $32.03^{\circ}$ & $7.97^{\circ}$\\ 
         \hline         
         1200 MHz& $30.95^{\circ}$  & $9.05^{\circ}$\\
         \hline
         1225 MHz& $29.89^{\circ}$  & $10.11^{\circ}$\\
         \hline         
         1250 MHz& $28.83^{\circ}$  & $11.17^{\circ}$\\
         \hline   
    \end{tabular}}
    \label{tab:1}
\end{table}
The difference in the beamformed angles, i.e., beam squint error caused by the wide bandwidth signals for the steering angle of $40^{\circ}$ when the phase shift only beamformer is used, is listed in Table \ref{tab:1}. It can be seen from the table that with the conventional phase shift-only technique, for a centre frequency of 1 GHz, the signals are steered to the required $39.99^{\circ}$. When the centre frequency is changed to 950 MHz and 1050 MHz (bandwidth of 100 MHz), the squint is increased, and the deviation increases as the bandwidth increases. 
Fig. \ref{beam_squint} illustrates the beam squint error in the phased array radars for different wide bandwidth signals. Results are shown for steering angles of $20^{\circ}$, $40^{\circ}$, $60^{\circ}$ and $90^{\circ}$. Other results are not included, considering the space constraints. 
From the plots, it is visible that the squint deviation is more for steering angles away from the boresight (array normal). Whereas, beams are thickly visible for lower steering angles (closer to boresight) such as $20^{\circ}$, indicating smaller deviations.

\begin{table}[!htb]
    \centering
    \caption{The maximum beam squint at different steering angles when conventional phase array beamformer is used}
    \resizebox{0.45\textwidth}{!}{
    \renewcommand{\arraystretch}{1.2}
    \begin{tabular}{|c|c|c|c|}
    \hline
         \makecell{Range of Centre\\ Frequencies (MHz)} & \makecell{Steering\\Angle ($\theta$)}  & \makecell{Maximum\\Beam Squint}  &  \makecell{Centre freq\\at which Max\\Beam Squint\\occurs}  \\ \hline
         \multirow{9}{*}{$ 750 -  1250$} & $0^{\circ}$ & $0^{\circ}$ & NA\\ 
         \cline{2-4}
         & $10^{\circ}$ &  $2.54^{\circ}$ & 750MHz\\
         \cline{2-4}
         & $20^{\circ}$  & $5.32^{\circ}$  & 750MHz\\
         \cline{2-4}
         & $30^{\circ}$  & $8.69^{\circ}$  & 750MHz \\
         \cline{2-4}
         & $40^{\circ}$  & $13.47^{\circ}$  & 750MHz\\
         \cline{2-4}
         & $50^{\circ}$  & $23.26^{\circ}$  & 750MHz \\
         \cline{2-4}
         & $60^{\circ}$  & $30.00^{\circ}$  & \makecell{825MHz, 800MHz,\\775MHz, 750MHz}\\
         \cline{2-4}
         & $70^{\circ}$ & $25.19^{\circ}$  & 1250MHz \\
         \cline{2-4}
         & $80^{\circ}$ & $32.38^{\circ}$ & 1250MHz \\
         \cline{2-4}
         & $90^{\circ}$ & $41.4^{\circ}$ & 1250MHz \\
         \hline
    \end{tabular}}
    \label{tab:2}
\end{table}

Table \ref{tab:2} shows the maximum beam squint error that occurred using conventional phase array beamforming for different steering angles across the range of bandwidths under consideration.  It can be seen from the table that as the steering angle increases, beam squint error also increases, which can also be seen from Fig. \ref{beam_squint}. This calls for the need to incorporate an additional technique that has the potential to minimize this squint angle deviation. To accommodate this, the received signal is first multiplied by the baseband delay and then phase-shifted, as mentioned in Section II. 

The integer part of the baseband delay is provided with the help of a fixed 30-tap FIR filter. To find the optimum fractional delay for each steering angle for the entire range of signal bandwidths, the delay is varied between -0.275 and 0.275, and the delay that minimizes the total beam squint error is found. For instance, for the steering angle of $40^\circ$, the baseband delay is varied between this range for the centre frequencies under consideration, the total beam squint error across all these frequencies is calculated, and the delay at which the minimum total squint is obtained is taken as the optimum delay for the steering angle of $40^\circ$. The variable fractional delay filter is tuned for this delay, which is multiplied with the received signal as shown in Fig. \ref{fig4}. The resultant signal is then phase shifted by $f_c \tau_n=f_c (n-1) d\frac{ \sin \theta}{c}$ to recover the original signal with minimum beam squint error.

\begin{table}[!htb]
    \centering
    \caption{The beam squint removal for wide bandwidth signals when optimum baseband delay technique is combined with phase compensation for the steering angle of $40^\circ$}
    \resizebox{0.29\textwidth}{!}{
    \renewcommand{\arraystretch}{0.92}
    \begin{tabular}{|c|c|c|}
    \hline
         \makecell{Centre\\Frequencies} &\makecell{Steered\\Angle ($\theta'$)}  & \makecell{Beam Squint\\Error ($\Delta \phi$)}   \\
         \hline
         750 MHz& $40.04^{\circ}$  & $0.04^{\circ}$\\
         \hline
         775 MHz& $39.99^{\circ}$  & $0.01^{\circ}$\\
         \hline    
         800 MHz& $40.06^{\circ}$  & $0.06^{\circ}$\\
         \hline
         825 MHz& $40.02^{\circ}$  & $0.02^{\circ}$\\
         \hline        
         850 MHz& $40.08^{\circ}$  & $0.08^{\circ}$\\
         \hline
         875 MHz& $40.02^{\circ}$  & $0.02^{\circ}$\\
         \hline         
         900 MHz& $39.97^{\circ}$ & $0.03^{\circ}$\\
         \hline
         925 MHz& $40.06^{\circ}$  & $0.06^{\circ}$\\
         \hline         
         950 MHz& $40.02^{\circ}$  & $0.02^{\circ}$\\
         \hline
         975 MHz& $39.99^{\circ}$ & $0.01^{\circ}$\\
         \hline         
         1000 MHz & $39.32^{\circ}$ & $0.68^{\circ}$\\ 
         \hline
         1025 MHz & $40.02^{\circ}$ & $0.02^{\circ}$\\ 
         \hline         
         1050 MHz& $39.97^{\circ}$ & $0.03^{\circ}$\\
         \hline
         1075 MHz & $39.95^{\circ}$ & $0.05^{\circ}$\\ 
         \hline         
         1100 MHz & $40.04^{\circ}$ & $0.04^{\circ}$\\ 
         \hline
         1125 MHz & $39.99^{\circ}$ & $0.01^{\circ}$\\ 
         \hline         
         1150 MHz& $39.92^{\circ}$ & $0.08^{\circ}$\\
         \hline
         1175 MHz & $39.97^{\circ}$ & $0.03^{\circ}$\\ 
         \hline         
         1200 MHz& $39.92^{\circ}$ & $0.08^{\circ}$\\
         \hline
         1225 MHz& $39.99^{\circ}$ & $0.01^{\circ}$\\
         \hline         
         1250 MHz& $39.97^{\circ}$ & $0.03^{\circ}$\\
         \hline   
    \end{tabular}}
    \label{tab:3}
\end{table}

The benefit of the proposed optimal baseband delay-based architecture can be clearly understood from the results shown in Table \ref{tab:3} for a desired steering angle of $40^{\circ}$. The optimal fractional delay is found as 0.176s for the steering angle of $40^{\circ}$, and the beam squint error at this delay value for different centre frequencies is listed in this table. The comparison of the beam squint values in Tables \ref{tab:1} and  \ref{tab:3} clearly shows the efficacy of the proposed technique in time delay compensation and squint reduction for wideband signals. The steered angle for the same set of centre frequencies is very close to the desired steering angle, resulting in a very reduced squint compared to phase shift-only beamforming.

\begin{figure}
    \centering
    \subfigure{\includegraphics[width=0.22\textwidth, keepaspectratio]{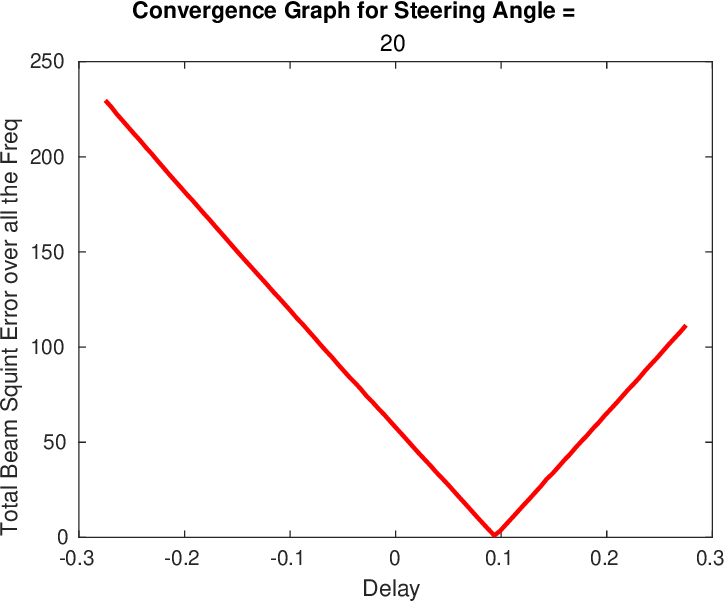}}
    \subfigure{\includegraphics[width=0.22\textwidth, keepaspectratio]{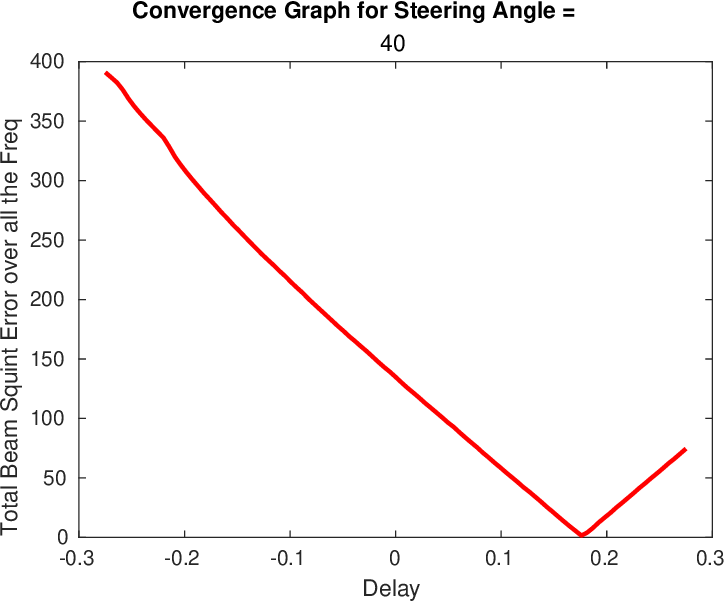}}
    \subfigure{\includegraphics[width=0.22\textwidth, keepaspectratio]{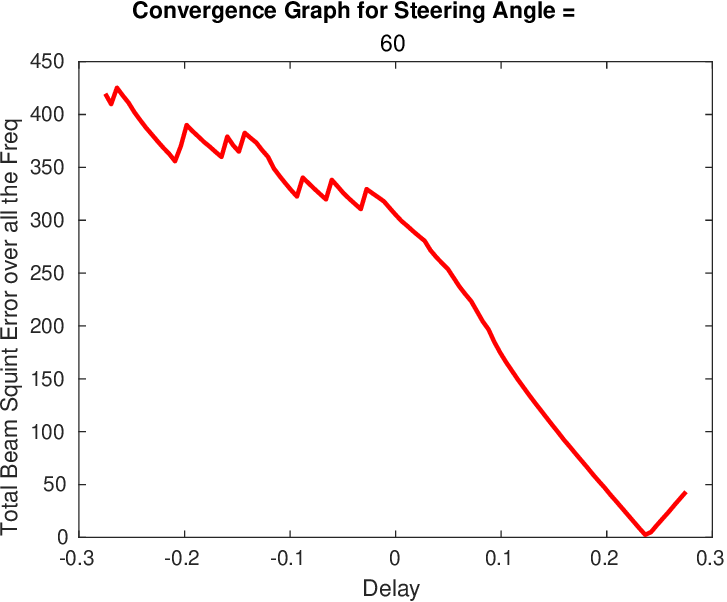}}
    \subfigure{\includegraphics[width=0.22\textwidth, keepaspectratio]{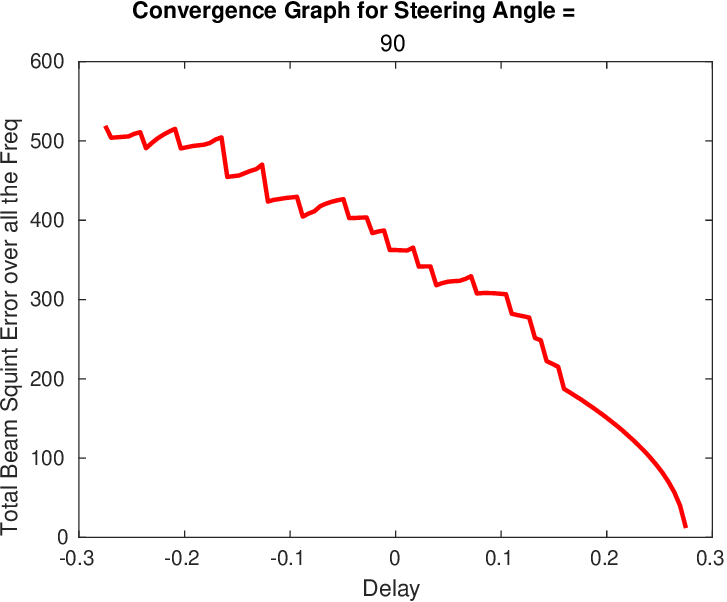}}
    \caption{Convergence Graph for Steering Angles (a) $20^{\circ}$ (b) $40^{\circ}$ (c) $60^{\circ}$ (d) $90^{\circ}$}
    \label{fig:conv_graph}
\end{figure}
\begin{table}[!htb]
    \centering
    \caption{The beam squint removal for wide bandwidth signals when optimum baseband delay technique is combined with phase compensation for various steering angles}
    \resizebox{0.45\textwidth}{!}{
    \renewcommand{\arraystretch}{1.1}
    \begin{tabular}{|c|c|c|c|c|}
    \hline
         \makecell{Range of\\Centre\\Frequencies \\(MHz)} & \makecell{Steering\\Angle ($\theta$)} &\makecell{Optimum\\Delay ($s$)} & \makecell{Maximum\\Beam Squint} &  \makecell{Centre freq\\at which max\\beam squint\\occurs} \\ \hline
         \multirow{9}{*}{\makecell{750 - 1250}} & $0^{\circ}$ & 0 & $0^{\circ}$ & NA\\ 
         \cline{2-5}
         & $10^{\circ}$ & 0.0495 & $0.21^{\circ}$ & \makecell{1000 MHz}\\
         \cline{2-5}
         & $20^{\circ}$ & 0.0935 & $0.42^{\circ}$ & 1000 MHz\\
         \cline{2-5}
         & $30^{\circ}$ & 0.1375 & $0.59^{\circ}$ & 1000 MHz\\
         \cline{2-5}
         & $40^{\circ}$ & 0.1760 & $0.68^{\circ}$ & 1000 MHz\\
         \cline{2-5}
         & $50^{\circ}$ & 0.2090 & $0.67^{\circ}$ & 1000 MHz\\
         \cline{2-5}
         & $60^{\circ}$  & 0.2365 & $0.56^{\circ}$& 1000 MHz\\
         \cline{2-5}
         & $70^{\circ}$ & 0.2585 & $0.39^{\circ}$ & 1000 MHz\\
         \cline{2-5}
         & $80^{\circ}$ & 0.2695 & $0.53^{\circ}$ & 850 MHz\\
         \cline{2-5}
         & $90^{\circ}$ & 0.2750 & $2.25^{\circ}$ & 1150 MHz\\
         \hline
    \end{tabular}}
    \label{tab:4}
\end{table}
\begin{figure}[!htb]
	\centering
	\subfigure{\includegraphics[width=0.22\textwidth, keepaspectratio]{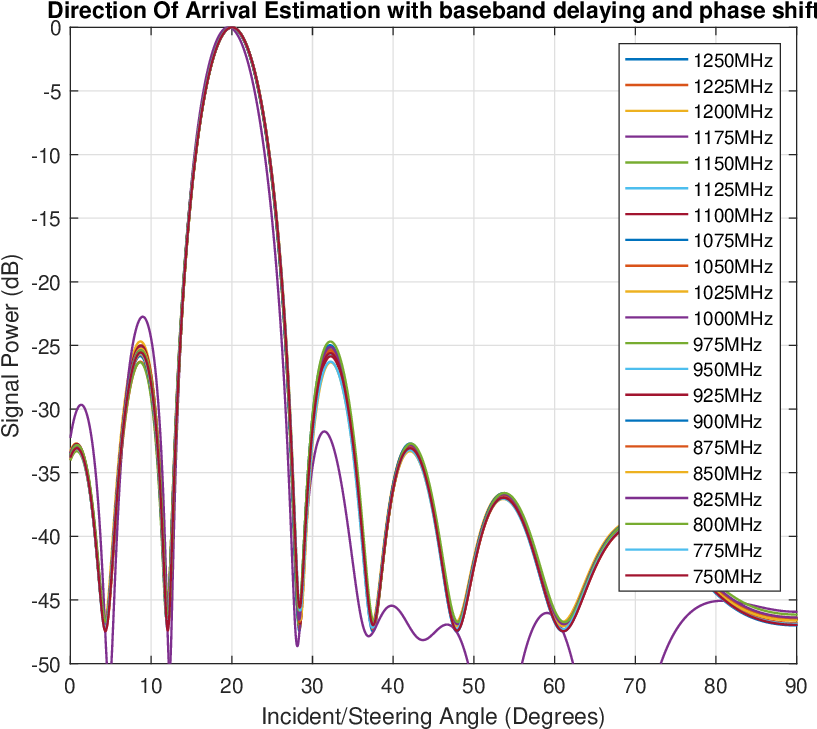}}
	\subfigure{\includegraphics[width=0.22\textwidth, keepaspectratio]{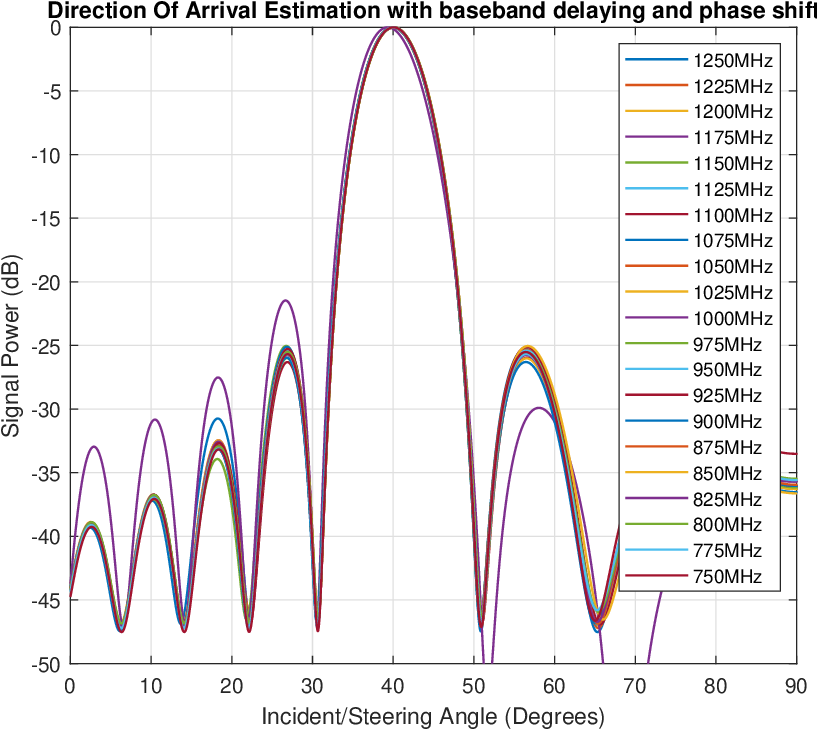}}
	\subfigure{\includegraphics[width=0.22\textwidth, keepaspectratio]{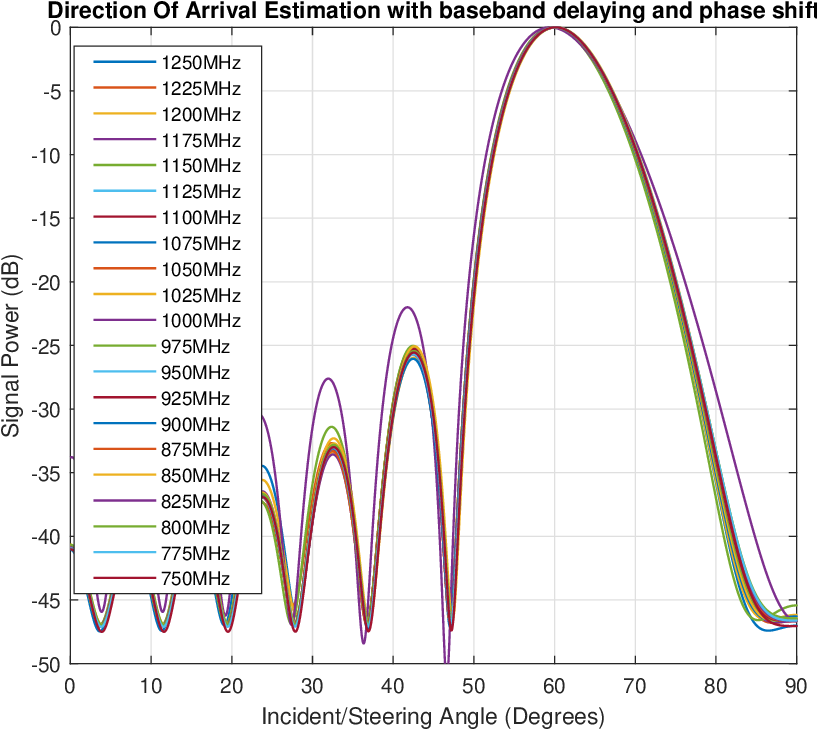}}
	\subfigure{\includegraphics[width=0.22\textwidth, keepaspectratio]{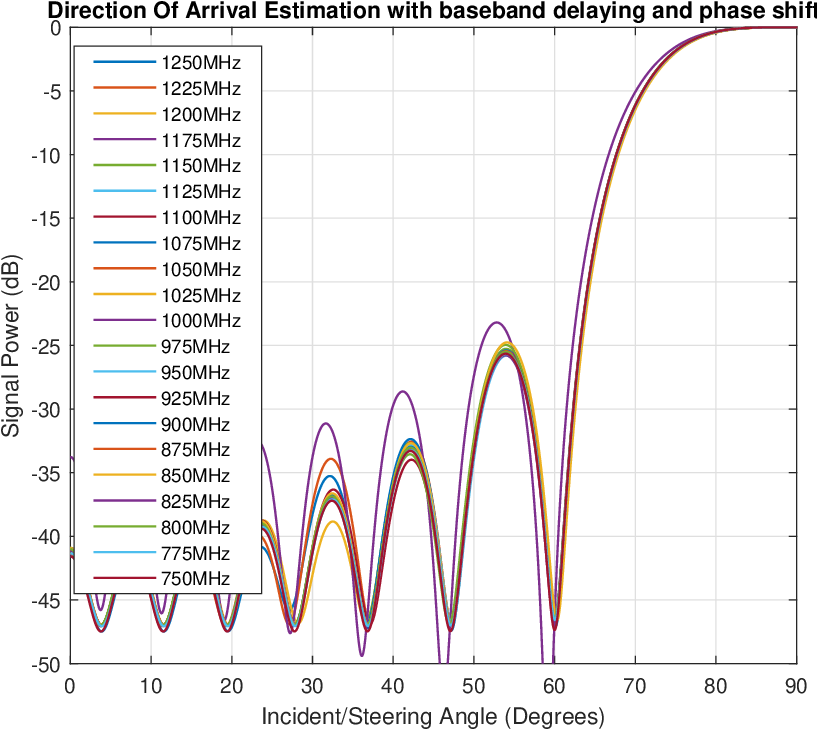}}
	\caption{\small Beam Squint Error in Phase array Radars for Wide Bandwidth Signals when steering angle is (a) $20^{\circ}$ (b) $40^{\circ}$ (c) $60^{\circ}$(d) $90^{\circ}$}
	\label{beamsqnt_removal}%
\end{figure}
A visual representation that shows how the total beam squint error values converge as delay varies is shown in Fig. \ref{fig:conv_graph} for steering angles $20^{\circ}$, $40^{\circ}$, $60^{\circ}$ and $90^{\circ}$. The Y-axis represents the total beam squint error and the X-axis represents the delay. 
Convergence happens at higher delay values for higher steering angles and vice versa.
For the set of frequencies under consideration, the optimum delay obtained for each steering angle in the range of $0^{\circ}$ to $90^{\circ}$ is given in Table \ref{tab:4}. 
The maximum squint error and centre frequency at which it is obtained are also listed in the table. The maximum beam squint obtained is within the limit of the design specification of $1^{0}$ for all the bandwidths and steering angles under consideration except at $90^{0}$. At $90^{0}$, the beam squint slightly exceeds the required specification for a few frequencies, such as 750 MHz, 800 MHz, 825 MHz, 975 MHz, and 1150 MHz.

The beam pattern plots with squint eliminated results for the steering angles  $20^{\circ}$, $40^{\circ}$, $60^{\circ}$, $90^{\circ}$ are shown in Fig. \ref{beamsqnt_removal}. Thus, it can be seen that the beam squint error is reduced to minimal values for the entire frequency range for all the steering angles under consideration.\\
 Table \ref{tab:5} provides a performance comparison between this work and contemporary state-of-the-art works. The digital beamformer in this work has 16 elements. It achieves baseband true-time delay, which ranges from 0 to 7.5 ns as in [14] but differs in bandwidth range, which here is considered as 25–250 MHz. The beam squint error obtained is as small as 0.68\degree, except for a slight rise in the case of 90\degree.
\begin{table}[h]
\centering
    \caption{Comparison with related papers in literature}
    \resizebox{0.47\textwidth}{!}{
    \renewcommand{\arraystretch}{1.1}
\begin{tabular}{|l|c|c|c|c|}
\hline
\textbf{Work for comparison}                                                   & \textbf{This work}                                        & \textbf{[20]}                                        & \textbf{[21]}                                         & \textbf{{[}14{]}}                                         \\ \hline
\textbf{\begin{tabular}[c]{@{}l@{}}Delay Implementation\\ Method\end{tabular}} & \begin{tabular}[c]{@{}c@{}}True Time\\ Delay\end{tabular} & \begin{tabular}[c]{@{}c@{}}Phase\\ Shifting\end{tabular} & \begin{tabular}[c]{@{}c@{}}Phase\\ Shifting\end{tabular} & \begin{tabular}[c]{@{}c@{}}True Time\\ Delay\end{tabular} \\ \hline
\textbf{\begin{tabular}[c]{@{}l@{}}Number of Antenna\\  elements\end{tabular}} & 16                                                        & 8                                                        & 16                                                       & 16                                                        \\ \hline
\textbf{Bandwidth (MHz)}                                                       & 25-250                                                    & 20                                                       & 100                                                      & 100                                                       \\ \hline
\textbf{Time delay range (ns)}                                                 & 0-7.5                                                    & -                                                        & -                                                        & 0-7.5                                                    \\ \hline
\textbf{Beam squint error (\degree)}                            & \textless{}1                                              & \textless{}16                                            & \textless{}18                                            & \textless{}1                                              \\ \hline
\end{tabular}}
\label{tab:5}
\end{table}
\section{Conclusion}
This paper extensively studies the beam squint issues when dealing with wideband signals in phased array radar receivers. An effective solution for this problem is proposed through the use of fine-tunable baseband delaying technique along with the phase shifters. A wide range of signal frequencies up to $\pm250$MHz with a carrier frequency of 1GHz  and steering angles spanning from $0^{\circ}$ to $90^{\circ}$ are considered for the analysis. The proposed scheme of combining integer baseband delay and fractional baseband delay with phase shift has proved to be an efficient way of removing beam squint errors when dealing with wide bandwidth signals.
The findings of the simulation validate the viability of the proposed approach for wide-bandwidth signals across a broad range of steering angles.  The demonstrated functionality across a broad range of parameters suggests the potential applicability of this technique in diverse radar scenarios, emphasizing its significance for enhancing the accuracy and reliability of target detection systems.

\end{document}